\documentclass[amstex,12pt,thmsa,sw20bams]{article}%
\usepackage{amsfonts}
\usepackage[utf8]{inputenc}
\usepackage{graphicx}
\usepackage[francais]{babel}%
\usepackage{amsmath}%
\usepackage{amssymb}
\providecommand{\U}[1]{\protect\rule{.1in}{.1in}}
\ifx\pdfoutput\relax\let\pdfoutput=\undefined\fi
\newcount\msipdfoutput
\ifx\pdfoutput\undefined\else
\ifcase\pdfoutput\else
\ifx\paperwidth\undefined\else
\ifdim\paperheight=0pt\relax\else\pdfpageheight\paperheight\fi
\ifdim\paperwidth=0pt\relax\else\pdfpagewidth\paperwidth\fi
\fi\fi\fi
\begin{document}

\author{Choulakian V., Universit\'{e} de Moncton, Canada
\and vartan.choulakian@umoncton.ca}
\title{Some notes on Goodman's marginal-free correspondence analysis}
\date{February 2021}
\maketitle

\begin{abstract}
In his seminal paper Goodman (1996) introduced marginal-free correspondence
analysis; where his principal aim was to reconcile Pearson correlation measure
with Yule's association measure in the analysis of contingency tables. We show
that marginal-free correspondence analysis is a particular case of
correspondence analysis with prespecified weights studied in the beginning of
the 1980s by Benz\'{e}cri and his students. Furthermore, we show that it is
also a particular first-order approximation of logratio analysis with uniform weights.



\end{abstract}

\section{\textbf{Introduction}}

Correspondence analysis (CA) and logratio analysis (LRA) are two popular
methods for the analysis and visualization of a contingency table (two-way
frequency counts data having $I$ rows and $J$ columns) or a compositional data
set ($I$ individuals, also named samples, of $J$ compositional parts). The
reference book on CA is Benz\'{e}cri (1973); Beh and Lombardo (2014) present a
panoramic review of CA and its variants.

LRA includes two independently well developed methods: RC association models
for the analysis of contingency tables by Goodman (1979, 1981a, 1981b, 1991,
1996) and compositional data analysis (CoDA) by Aitchison (1986). CA and LRA
are based on three different principles: CA on Benz\'{e}cri's
\textit{distribututional equivalence principle}, RC association models on
Yule's \textit{scale invariance principle}, and CoDA on Aitchison's
\textit{subcompositional coherence principle}.

From a statistical point of view there is a fundamental difference between the
structures of a two-way contingency table $\mathbf{N}=(n_{ij})$ and a
compositional data set $\mathbf{X}=(x_{ij})$ for $i=1,...,I$ and $j=1,...,J$;
while from a mathematical point of view the form of the resulting statistical
equations arising from different departure assumptions may be identical in
Goodman's RC association models and Aitchison's CoDA.

Goodman (1996, equation (46)) in his seminal paper introduced marginal-free
correspondence analysis (mfCA); where his principal aim was to reconcile
Pearson correlation measure with Yule's association measure in the analysis of
contingency tables. In this paper, we show that mfCA is a particular case of
CA with prespecified weights, which has been studied in the beginning of 1980s
under the direction of Benz\'{e}cri. In Benz\'{e}cri's edited journal
\textit{Les Cahiers de l'Analyse des Donn\'{e}es}, the following papers
appeared [Madre (1980), Cholakian (1980, 1984), Benz\'{e}cri (1983a, 1983b),
Benz\'{e}cri et al. (1980) and Moussaoui (1987)]. Furthermore, we show that
mfCA is also a particular first-order approximation of LRA analysis with
uniform weights.

This paper is organized as follows: Section 2 presents three different basic
ways of representing the concept of interaction in a contingency table;
section 3 discusses the the important consequences of Yule's scale invariance
association index; section 4 presents Goodman's marginial-free CA; section 5
discusses an example; section 6 presents the R code to do the computations;
finally we conclude in section 7.

\section{Preliminaries on analysis of contingency tables}

Let $\mathbf{P=N/}n=(p_{ij})$ of size $I\times J$ be the associated
correspondence matrix (probability table) of a contingency table \textbf{N}.
We define as usual $p_{i+}=\sum_{j=1}^{J}p_{ij}$ , $p_{+j}=\sum_{i=1}%
^{I}p_{ij},$ the vector $\mathbf{r=(}p_{i+})\in%
\mathbb{R}
^{I},$ the vector $\mathbf{c=(}p_{+j})\in%
\mathbb{R}
^{J}$, and $\mathbf{M}_{I}=Diag(\mathbf{r})$ the diagonal matrix having
diagonal elements $p_{i+},$ and similarly $\mathbf{M}_{J}=Diag(\mathbf{c}).$
We suppose that $\mathbf{M}_{I}$ and $\mathbf{M}_{J}$ are positive definite
metric matrices of size $I\times I$ and $J\times J$, respectively; this means
that the diagonal elements of $\mathbf{M}_{I}$ and $\mathbf{M}_{J}$ are
strictly positive.

\subsection{Independence of the row and column categories}

a) The $I$ row categories are independent of the $J$ column categories,
\begin{equation}
\sigma_{ij}=p_{ij}-p_{i+}p_{+j}=0,
\end{equation}
where $\sigma_{ij}$ is the residual matrix of $p_{ij}$ with respect to the
independence model $p_{i+}p_{+j}.$

\textbf{Remark 1}: The contingency table $\mathbf{N}=(n_{ij})$ can also be
represented (coded) as an indicator matrix $\mathbf{Z=}\left[  \mathbf{Z}%
^{I}\ \ \mathbf{Z}^{J}\right]  =\left[  \mathbf{(}z_{\alpha i})\ \ \mathbf{(}%
z_{\alpha j})\right]  \ $of size $n$ by $I+J,$ where $z_{\alpha i}=0$ if
individual $\alpha$ does not have level $i$ of the row variable, $z_{\alpha
i}=1$ if individual $\alpha$ has level $i$ of the row variable; $z_{\alpha
j}=0$ if individual $\alpha$ does not have level $j$ of the column variable,
$z_{\alpha j}=1$ if individual $\alpha$ has level $j$ of the column variable.
Note that $\mathbf{N}=(\mathbf{Z}^{I})^{\prime}\mathbf{Z}^{J}\ $and
$\sigma_{ij}=p_{ij}-p_{i+}p_{+j}$ is the covariance between the $i$-th column
of $\mathbf{Z}^{I}$ and the $j$-th column of $\mathbf{Z}^{J}$.

b) The independence assumption $\sigma_{ij}=0$ can also be interpreted in
another way as%

\begin{align}
\Delta_{ij}  & =(\frac{p_{ij}}{p_{i+}p_{+j}}-1)=0\\
& =\frac{1}{p_{i+}}(\frac{p_{ij}}{p_{+j}}-p_{i+})=0\nonumber\\
& =\frac{1}{p_{+j}}(\frac{p_{ij}}{p_{i+}}-p_{+j});\nonumber
\end{align}
this is the column and row homogeneity models. Benz\'{e}cri (1973, p.31) named
the conditional probability vector ($\frac{p_{ij}}{p_{+j}}$ for $i=1,...,I$
and $j$ fixed) the profile of the $j$th column; and the element $\frac{p_{ij}%
}{p_{i+}p_{+j}}$ the density function of the probability measure $(p_{ij})$
with respect to the product measure $p_{i+}p_{+j}$. The element $\frac{p_{ij}%
}{p_{i+}p_{+j}}$is named Pearson ratio in Goodman (1996) and Beh and Lombardo
(2014, p.123).

c) A third way to represent the indepence assumption $\sigma_{ij}=0$ and the
row and column homogeneity models $\Delta_{ij}=0$ is via the ($w_{i}^{R} $,
$w_{j}^{C})$ weighted loglinear formulation, equation (3), assuming $p_{ij}>0$
and defining $G_{ij}=\log(p_{ij}),$%

\begin{align}
\lambda_{ij}  & =0\\
& =G_{ij}-G_{i+}-G_{+j}+G_{++},\nonumber
\end{align}
where $G_{i+}=\sum_{j=1}^{J}G_{ij}w_{j}^{C},$ $G_{+j}=\sum_{i=1}^{I}%
G_{ij}w_{i}^{R}$ and $G_{++}=\sum_{j=1}^{J}\sum_{i=1}^{I}G_{ij}w_{j}^{C}%
w_{i}^{R}$; $w_{j}^{C}>0$ and $w_{i}^{R}>0,$ satisfying $\sum_{j=1}^{J}%
w_{j}^{C}=\sum_{i=1}^{I}w_{i}^{R}\ =1,$ are a priori fixed probability
weights. Two popular weights are marginal ($w_{j}^{C}=p_{+j}$, $w_{i}%
^{R}=p_{i+})$ and uniform ($w_{j}^{C}=1/J$,$w_{i}^{R}=1/I).$ This is implicit
in equation 7 in Goodman (1996) or equation 2.2.6 in Goodman (1991); and
explicit in Egozcue et al. (2015).

Equation (3) is equivalent to the logratios%
\[
\log(\frac{p_{ij}p_{i_{1}j_{1}}}{p_{ij_{1}}p_{i_{1}j}})=0\text{ \ \ for }i\neq
i_{1}\text{ and }j=j_{1},
\]
which Goodman (1979, equation 2.2) names it \textquotedblright null
association\textquotedblright\ model.

Equation (3) is also equivalent to%
\[
p_{ij}=\frac{\exp(G_{i+})\exp(G_{+j})}{\exp(G_{++})},
\]
from which we deduce that : under the independence assumption the marginal row
probability vector ($p_{i+})$ is proportional to the vector of weighted
geometric means ($\exp(G_{i+}));$ and a similar property is true also for the
columns; see for instance Egozcue et al. (2015).

\subsection{Interaction factorization}

Suppose the independence-homogeneity-null association models are not true,
then each of the three equivalent model formulations (1,2,3) can be
generalized to explain the nonindependence-nonhomogeneity-association, named
interaction, among the $I$ rows and the $J$ columns by adding $k$ bilinear
terms, where $k=rank(\mathbf{N)-}1$. We designate any one of the interaction
indices (1,2,3) by $\tau_{ij}.$

Benz\'{e}cri (1973, Vol.1, p. 31-32) emphasized the importance of row and
column weights or metrics in multidimensional data analysis; this is the
reason in the french data analysis circles any study starts with a triplet
$\mathbf{(X,M}_{I},\mathbf{M}_{J}\mathbf{)}$, where \textbf{X} represents the
data set, $\mathbf{M}_{I}=(Diag(m_{i}^{r}))$ is the metric defined on the rows
and $\mathbf{M}_{J}=(Diag(m_{j}^{c}))$ the metric defined on the columns. We
follow the same procedure where:

a) In covariance analysis, $\mathbf{X}=(\tau_{ij}\mathbf{)=(}\sigma_{ij})$ and
$(\mathbf{M}_{I},\mathbf{M}_{J})=(Diag(1/I),Diag(1/J));$

b) In CA, $\mathbf{X}=(\tau_{ij}\mathbf{)=(}\Delta_{ij})$ and $(\mathbf{M}%
_{I},\mathbf{M}_{J})=(Diag(p_{i+}),Diag(p_{+j}));$

c) In LRA, $\mathbf{X}=(\tau_{ij}\mathbf{)=(}\lambda_{ij})$ and $(\mathbf{M}%
_{I},\mathbf{M}_{J})=(Diag(w_{i}^{R}),Diag(w_{j}^{C}))$ with $\sum_{j=1}%
^{J}w_{j}^{C}=\sum_{i=1}^{I}w_{i}^{R}=1.$

We factorize the interactions in (1,2,3) by singular value decomposition (SVD)
or taxicab SVD (TSVD) as%
\begin{equation}
\tau_{ij}=\sum_{\alpha=1}^{k}f_{\alpha}(i)g_{\alpha}(j)/\delta_{\alpha}.
\end{equation}%
\[
\]
In the SVD case the parameters $(f_{\alpha}(i),g_{\alpha}(j),\delta_{\alpha})$
satisfy the conditions: for $\alpha,\beta=1,...,k$
\[
\delta_{\alpha}^{2}=\sum_{\alpha=1}^{k}f_{\alpha}^{\ \ 2}(i)m_{i}^{r}%
=\sum_{\alpha=1}^{k}g_{\alpha}^{2}(j)m_{j}^{c}
\]%
\[
0=\sum_{\alpha=1}^{k}f_{\alpha}(i)m_{i}^{r}=\sum_{\alpha=1}^{k}g_{\alpha
}(j)m_{j}^{c}
\]%
\[
0=\sum_{\alpha=1}^{k}f_{\alpha}(i)f_{\beta}(i)m_{i}^{r}=\sum_{\alpha=1}%
^{k}g_{\alpha}(j)g_{\beta}(j)m_{j}^{c}
\]
In the TSVD case the parameters $(f_{\alpha}(i),g_{\alpha}(j),\delta_{\alpha
})$ satisfy the conditions: for $\alpha,\beta=1,...,k$
\[
\delta_{\alpha}=\sum_{\alpha=1}^{k}|f_{\alpha}^{\ }|(i)m_{i}^{r}=\sum
_{\alpha=1}^{k}|g_{\alpha}(j)|m_{j}^{c}
\]%
\[
0=\sum_{\alpha=1}^{k}f_{\alpha}(i)m_{i}^{r}=\sum_{\alpha=1}^{k}g_{\alpha
}(j)m_{j}^{c}
\]%
\[
0=\sum_{\alpha=1}^{k}f_{\alpha}(i)\ sign(f_{\beta}(i))m_{i}^{r}=\sum
_{\alpha=1}^{k}g_{\alpha}(j)\ sign(g_{\beta}(j))\ m_{j}^{c}\text{\ \ for
}\alpha>\beta.
\]
A description of TSVD can be found, among others, in Choulakian (2006,
2016).\bigskip

\textbf{Remark 2}

a) In the case $(\tau_{ij}\mathbf{)=(}\sigma_{ij})$, the bilinear
decomposition (4) is also named interbattery analysis first proposed by Tucker
(1958); later on, Tenenhaus and Augendre (1996) reintroduced it within
correspondence analysis circles, where they showed that the Tucker
decomposition by SVD produced on some correpodence tables more interesting
structure, more interpretable, than CA.

b) In the case $(\tau_{ij}\mathbf{)=(}\Delta_{ij})$, the CA decomposition has
many interpretations. Essentially, for data analysis purposes Benz\'{e}cri
(1973) interpreted it as weighted principal components analysis of row and
column profiles. Another useful interpretation of CA, comparable to Tucker
interbattery analysis, is Hotelling(1936)'s canonical correlation analysis,
see Lancaster (1958) and Goodman (1991, 1996).

\section{Yule's principle of scale invariance}

We start by quoting Goodman (1996, section 10) to really understand Yule's
principle of scale invariance: \textquotedblright Pearson's approach to the
analysis of cross-classified data was based primarily on the bivariate normal.
He assumed that the row and column classifications arise from underlying
continuous random variables having a bivariate normal distribution, so that
the sample contingency table comes from a discretized bivariate normal; and he
then was concerned with the estimation of the correlation coefficient for the
underlying bivariate normal. On the other hand, Yule felt that, for many kinds
of contingency tables, it was not desirable in scientific work to introduce
assumptions about an underlying bivariate normal in the analysis of these
tables; and for such tables, he used, to a great extent, coefficients based on
the odds-ratios (for example, Yule's Q and Y), coefficients that did not
require any assumptions about underlying distributions. The Pearson approach
and the Yule approach appear to be wholly different, but a kind of
reconciliation of the two perspectives was obtained in Goodman
(1981a)\textquotedblright. An elementary exposition of these ideas with
examples can also be found in Mosteller (1968).

In the notation of our paper, Goodman's reconciliation is based on defining
the a priori weights in the association index (3), $\lambda_{ij}%
=\lambda(p_{ij},w_{j}^{C},w_{i}^{R}),$ where by its decomposition into
bilinear terms, mwLRA will correspond to Pearson's approach, while uwLRA to
Yule's approach. Because log-odds%
\begin{align}
\log(\frac{p_{ij}p_{i_{1}j_{1}}}{p_{ij_{1}}p_{i_{1}j}})  & =\lambda
_{ij}+\lambda_{_{i_{1}j_{1}}}-\lambda_{_{i_{1}j}}-\lambda_{ij_{1}}\nonumber\\
& =\sum_{\alpha=1}^{k}(f_{\alpha}(i)-f_{\alpha}(i_{1}))(g_{\alpha
}(j)-g_{\alpha}(j_{1})/\delta_{\alpha}.
\end{align}

To have a clear picture of LRA with general a priori prescribed weights
($w_{j}^{C},w_{i}^{R})$, we first study the properties of the association
index $\lambda_{ij},$ that distinguishes it from interaction indices (2,3).

\subsection{Scale invariance of an interaction index}

We are concerned with the property of scale dependence or independence of the
three interaction indices (1,2,3). We note that in (1,2,3), $p_{ij}$ depends
on $n_{ij},$ $p_{ij}=n_{ij}/\sum_{i,j}n_{ij}.$ To emphasize this dependence,
we express an interaction index by $\tau_{ij}(n_{ij})=\tau(p_{ij},m_{i}%
^{R},m_{j}^{C})$ where: in the case of the association index $\tau_{ij}%
(n_{ij})=\lambda_{ij}$ is defined in (3), in the case of the nonhomogeneity
index $\tau_{ij}(n_{ij})=\Delta_{ij}$ is defined in (2), and in the case of
the nonindependence index $\tau_{ij}(n_{ij})=\sigma_{ij} $ is defined in (1).
Following Yule (1912), we state the following\bigskip

\textbf{Definition 1}: An interaction index $\tau_{ij}(n_{ij})$ is scale
invariant if $\tau_{ij}(n_{ij})=\tau_{ij}(a_{i}n_{ij}b_{j})$ for scales
$a_{i}>0$ and $b_{j}>0$.\bigskip

It is important to note that Yule's principle of scale invariance concerns a
function of four interaction terms, see equation (5); while in Definition 1
the invariance concerns each interaction term.

It is evident that the interaction indices (1 and 2) are not scale invariant:
because they are marginal-dependent.

Concerning the association index (3) we have\bigskip

\textbf{Lemma 1}: The association index (3) is scale invariant.

\textit{Proof}: Let $n^{\ast}=\sum_{i,j}a_{i}n_{ij}b_{j},$ then
\begin{align}
\tau_{ij}(a_{i}n_{ij}b_{j})  & =\lambda(a_{i}n_{ij}b_{j}/n^{\ast},w_{j}%
^{C},w_{i}^{R})\nonumber\\
& =\log(a_{i}n_{ij}b_{j}/n^{\ast})-\sum_{j=1}^{J}w_{j}^{C}\log(a_{i}%
n_{ij}b_{j}/n^{\ast})\nonumber\\
& -\sum_{i=1}^{I}w_{i}^{R}\log(a_{i}n_{ij}b_{j}/n^{\ast})+\sum_{j=1}^{J}%
\sum_{i=1}^{I}w_{j}^{C}w_{i}^{R}\log(a_{i}n_{ij}b_{j}/n^{\ast})\nonumber\\
& =\lambda(n_{ij},w_{j}^{C},w_{i}^{R})=\lambda(p_{ij},w_{j}^{C},w_{i}%
^{R})=\tau_{ij}(n_{ij})\nonumber\\
& =\lambda(a_{i}p_{ij}b_{j},w_{j}^{C},w_{i}^{R}).
\end{align}

\textbf{Lemma} 2: To a first-order approximation, $\lambda_{ij}\approx
(\frac{p_{ij}}{w_{j}^{C}w_{i}^{R}}-\frac{p_{i+}}{w_{i}^{R}}-\frac{p_{+j}%
}{w_{j}^{C}}+1).$

\textit{Proof}: The average value of the density function $\frac{p_{ij}}%
{w_{j}^{C}w_{i}^{R}}$ with respect to the product measure $w_{j}^{C}w_{i}^{R}$
is 1; so the $IJ$ values $\frac{p_{ij}}{w_{j}^{C}w_{i}^{R}}$ are distributed
around 1. By Taylor series expansion of $\log x$ in the neighborhood of $x=1$,
we have to a first-order $\log x\approx x-1.$ Putting $a_{i}=1/w_{i}^{R}$ and
$b_{j}=1/w_{j}^{C}$ in (6), and by using $\log(\frac{p_{ij}}{w_{j}^{C}%
w_{i}^{R}})\approx\frac{p_{ij}}{w_{j}^{C}w_{i}^{R}}-1$%
\begin{align*}
\lambda(p_{ij},w_{j}^{C},w_{i}^{R})  & =\lambda(\frac{p_{ij}}{w_{j}^{C}%
w_{i}^{R}},w_{j}^{C},w_{i}^{R})\\
& =\log(\frac{p_{ij}}{w_{j}^{C}w_{i}^{R}})-\sum_{j=1}^{J}w_{j}^{C}\log
(\frac{p_{ij}}{w_{j}^{C}w_{i}^{R}})\\
& -\sum_{i=1}^{I}w_{i}^{R}\log(\frac{p_{ij}}{w_{j}^{C}w_{i}^{R}})+\sum
_{j=1}^{J}\sum_{i=1}^{I}w_{j}^{C}w_{i}^{R}\log(\frac{p_{ij}}{w_{j}^{C}%
w_{i}^{R}}\\
& \approx\frac{p_{ij}}{w_{j}^{C}w_{i}^{R}}-1-(\frac{p_{i+}}{w_{i}^{R}%
}-1)-(\frac{p_{+j}}{w_{j}^{C}}-1)+0,
\end{align*}
which is the required result.\bigskip

\textbf{Remark 3}: Lemma 2 provides a first order approximation to mwTLRA and
uwTLRA, where we see that both first-order approximations are
marginal-dependent but in different ways.

a) In the case $(a_{i},b_{j})=(1/p_{i+},1/p_{+j})$ and$\ (w_{j}^{C},w_{i}%
^{R})=(p_{+j},p_{i+})$ in Lemma 2, $\lambda_{ij}=\lambda(p_{ij},p_{+j}%
,p_{i+})\approx$ $\frac{p_{ij}}{p_{+j}p_{i+}}-1;$ which implies that CA (or
TCA) is a first-order approximation of mwLRA (or mwTLRA), a result stated in
Cuadras et al. (2006).

b) In the case $(a_{i},b_{j})=(I,J)$ and$\ (w_{j}^{C},w_{i}^{R})=(1/I,1/J)$ in
Lemma 2, $\lambda_{ij}=\lambda(p_{ij},1/J,1/I)\approx$ $IJp_{ij}%
-Ip_{i+}-Jp_{+j}+1;$ which implies that the bilinear expansion of the right
side by TSVD (or SVD) is a first-order approximation of uwTLRA (or uwLRA).

In this subsection, we discussed the approximation of LRA to CA related
methods. Greenacre (2009) posed the reciprocal question: when CA related
methods converge to LRA? And he stated two results; which we provide a proof
in the following subsection.

\subsection{Box-Cox transformation}

Theoretically CA and LRA have been presented in a unified mathematical
framework via Box-Cox transformation by Goodman (1996), where the bilinear
terms have been estimated by SVD. Goodman's framework was further considered,
among others, by Cuadras et al. (2006), Greenacre (2009, 2010), and Cuadras
and Cuadras (2015 ).

Consider the triplet \textbf{(X, Q, D)}, where $\mathbf{X=(}x_{ij})$ with
$x_{ij}>0$ represents the data set, and $(\mathbf{Q,D})=(Diag(w_{i}%
^{R}),Diag((w_{j}^{C}))$ with $\sum_{j=1}^{J}w_{j}^{C}=\sum_{i=1}^{I}w_{i}%
^{R}=1.$ Let $\alpha$ be a nonnegative real number. Following Goodman (1996,
equations (3,4,5)), we define the interaction index,
\begin{align}
Int(\frac{x_{ij}^{\alpha}}{\alpha},w_{j}^{C},w_{i}^{R})  & =\frac
{x_{ij}^{\alpha}}{\alpha}-\sum_{j=1}^{J}w_{j}^{C}\frac{x_{ij}^{\alpha}}%
{\alpha}-\sum_{i=1}^{I}w_{i}^{R}\frac{x_{ij}^{\alpha}}{\alpha}+\sum_{j=1}%
^{J}\sum_{i=1}^{I}w_{j}^{C}w_{i}^{R}\frac{x_{ij}^{\alpha}}{\alpha}\nonumber\\
& =(\frac{x_{ij}^{\alpha}-1}{\alpha})-\sum_{j=1}^{J}w_{j}^{C}(\frac
{x_{ij}^{\alpha}-1}{\alpha})\nonumber\\
& -\sum_{i=1}^{I}w_{i}^{R}(\frac{x_{ij}^{\alpha}-1}{\alpha})+\sum_{j=1}%
^{J}\sum_{i=1}^{I}w_{j}^{C}w_{i}^{R}(\frac{x_{ij}^{\alpha}-1}{\alpha}).\\
& \nonumber
\end{align}
Using the well-known result based on Hopital's rule, lim$_{\alpha\rightarrow
0}(\frac{x_{ij}^{\alpha}-1}{\alpha})=\log(x_{ij}),$ (7) converges to
\begin{equation}
\lambda(x_{ij},w_{j}^{C},w_{i}^{R})=\log(x_{ij})-\sum_{j=1}^{J}w_{j}^{C}%
\log(x_{ij})-\sum_{i=1}^{I}w_{i}^{R}\log(x_{ij})+\sum_{j=1}^{J}\sum_{i=1}%
^{I}w_{j}^{C}w_{i}^{R}\log(x_{ij}).
\end{equation}
We consider two cases of (7, 8):

a) $\lambda(x_{ij},w_{j}^{C},w_{i}^{R})=\lambda(p_{ij},p_{+j},p_{i+}%
)=\lambda(p_{ij}/(p_{+j}p_{i+}),p_{+j},p_{i+}),$ which is the interaction term
of mwLRA, and equivalent to Result 2 in Greenacre (2010).

b)\ \ $\lambda(x_{ij},w_{j}^{C},w_{i}^{R})=\lambda(p_{ij},1/J,1/I)=\lambda
(IJp_{ij},1/J,1/I)=\lambda(p_{ij}/(p_{+j}p_{i+}),1/J,1/I),$ which is the
interaction term of uwLRA; this is similar to Result 1 in Greenacre (2010).

Equation (7) can also be applied differently, where:

In (7) we replace $w_{j}^{C}$ by$\ w_{j}^{C}(\alpha)=\frac{\sum_{i=1}%
^{I}x_{ij}^{\alpha}}{\sum_{i=1}^{I}\sum_{j=1}^{J}x_{ij}^{\alpha}},\ w_{i}^{R}$
by$\ w_{i}^{R}(\alpha)=\frac{\sum_{j=1}^{J}x_{ij}^{\alpha}}{\sum_{i=1}^{I}%
\sum_{j=1}^{J}x_{ij}^{\alpha}};$ we see that lim$_{\alpha\rightarrow0}%
w_{j}^{C}(\alpha)=\frac{I}{IJ}=1/J;$ similarly lim$_{\alpha\rightarrow0}%
w_{i}^{R}(\alpha)=1/I.$ Then we get%
\[
\text{lim}_{\alpha\rightarrow0}Int(x_{ij}^{\alpha},w_{j}^{C}(\alpha),w_{i}%
^{R}(\alpha))=\lambda(x_{ij},1/J,1/I).
\]
In particular lim$_{\alpha\rightarrow0}Int(x_{ij}^{\alpha},w_{j}^{C}%
(\alpha),w_{i}^{R}(\alpha))=\lambda(p_{ij},1/J,1/I)=\lambda(IJp_{ij}%
,1/J,1/I)=\lambda(p_{ij}/(p_{+j}p_{i+}),1/J,1/I)$.

\section{CA with prescribed weights and Goodman's mfCA}

CA with prescribed weights is done in two steps in the following way: We
observe a probability table $\mathbf{P}=(p_{ij})$ of size $I$ by $J$. Let
$\mathbf{Q}$ of size $I$ by $J$ be an \textit{unknown} probability table with
\textit{known} marginals $q_{i+}$ and $q_{+j}.$ The two steps are:

Step1: We construct \textbf{Q} which is in a sense \textquotedblright nearest
to $\mathbf{P}$\textquotedblright\textit{. }Two general criteria are:
$Int(q_{ij},q_{+j},q_{i+})=\lambda(q_{ij}=a_{i}p_{ij}b_{j},q_{+j},q_{i+})$
based on (3) and min$_{q_{ij}}\sum_{i,j}(\frac{q_{ij}}{q_{i+}q_{+j}}%
-\frac{p_{ij}}{p_{i+}p_{+j}})_{{}}^{2}q_{i+}q_{+j}$ based on (2).

Step 2: We apply CA to the constructed probability \textbf{Q}
\[
\frac{q_{ij}-q_{i+}q_{+j}}{q_{i+}q_{+j}}=\sum_{\alpha=1}^{k}f_{\alpha
}(i)g_{\alpha}(j)/\delta_{\alpha},
\]
which represents CA of $\mathbf{P}$ with prespecified weights ($q_{i+}%
,q_{+j})$. Cholakian (1980) presents an example, where both criteria have been
applied and similar results have been obtained.

In the particular case, where $Int(q_{ij},1/J,1/I)=\lambda(q_{ij}=a_{i}%
p_{ij}b_{j},1/J,1/I),$ we get Goodman's mfCA, see Goodman (1996, equation
(46)). $\mathbf{Q=}(q_{ij})$ is related to $\mathbf{P=(}p_{ij})$ via the
strictly positive scales ($a_{i},b_{j}),$ that keeps the association between
the $i$-th row and the $j$-th column unchanged. The famous iterative
proportional fitting algorithm (IPF) is used to construct \textbf{Q}. That is,
the constructed probability table ($q_{ij})$ has uniform marginals
$q_{+j}=1/J$ and $q_{i+}=1/I.$ So in Step 2, CA representation is%
\begin{equation}
IJq_{ij}-1=\sum_{\alpha=1}^{k}f_{\alpha}(i)g_{\alpha}(j)/\delta_{\alpha},
\end{equation}
which represents a first-order approximation to both uwLRA and mwLRA by Remark
3. Furthermore, by Remark 2 we see that mfCA can be interpreted both as Tucker
and Hotelling decompositions.

\section{Examples}

We present the analysis of two datasets for comparative purposes.

\subsection{Example 1}

This dataset is contrived and taken from Goodman (1991, Table 10(1), that we
reproduce below

$%
\begin{bmatrix}
4 & 10 & 1\\
10 & 50 & 10\\
1 & 10 & 4
\end{bmatrix}
$

According to Goodman, LRA has one principal dimension, while CA has 2
principal dimensions. Here we compare the dispersion results of the 4 methods:
CA, TCA, mfCA and mfTCA.

In CA: $\delta_{1}=corr(f_{1}(i),g_{1}(j)=0.20$ and $\delta_{2}=corr(f_{2}%
(i),g_{2}(j)=0.048$

In mfCA: $\delta_{1}=corr(f_{1}(i),g_{1}(j)=0.41$ and $\delta_{2}%
=corr(f_{2}(i),g_{2}(j)=0.050$

In TCA: $\ \ \ \ \delta_{1}=0.070$ and $\delta_{2}=0.034$

In mfTCA: $\delta_{1}=0.285$ and $\delta_{2}=0.043$

\subsection{Example 2}

We consider the \textit{rodent} data set of size 28 by 9 found in TaxicabCA in
R package. This is an abundance data set of 9 kinds of rats in 28 cities in
California. It can be considered both a contingency table and a compositional
data set. Choulakian (2017) analyzed it by comparing the CA and TCA maps;
furthermore Choulakian (2021) showed that it has quasi-2-blocks structure.
Here we compare the dispersion results of the first 2 principal dimensions in
the 4 methods: CA, TCA, mfCA and mfTCA:

In CA: $\delta_{1}=corr(f_{1}(i),g_{1}(j)=0.864$ and $\delta_{2}%
=corr(f_{2}(i),g_{2}(j)=0.678$

In mfCA:$\ \delta_{1}=corr(f_{1}(i),g_{1}(j)=0.827$ and $\delta_{2}%
=corr(f_{2}(i),g_{2}(j)=0.679$

In TCA$:\ \ \ \ \delta_{1}=0.478$ and $\delta_{2}=0.422$

In mfTCA: $\delta_{1}=0.743$ and $\delta_{2}=0.541$

The curious reader can apply the R code below to campare the 4 maps: CA, mfCA,
TCA and mfTCA.

\section{R code}

\#

\# install packages

install.packages(c("ipfr", "ca", "TaxicabCA"))

\#

library(TaxicabCA)

dataMatrix = as.matrix(rodent)

nRow
$<$%
- nrow(dataMatrix)

nCol
$<$%
- ncol(dataMatrix)

ssize
$<$%
- sum(dataMatrix)

\#

\#Computation of Q matrix of rodent

library(ipfr)

mtx
$<$%
- dataMatrix

row\_targets
$<$%
- rep(ssize/nRow, nRow)

column\_targets
$<$%
- rep(ssize/nCol, nCol)

QMatrix
$<$%
- ipu\_matrix(mtx, row\_targets, column\_targets)

rownames(QMatrix)
$<$%
- paste("",1:nRow,sep="")

colnames(QMatrix)
$<$%
- paste("C",1:nCol,sep="")

\#

\#CA map of rodent dataset

library(ca)

plot(ca(dataMatrix))

\#mfCA map of rodent

plot(ca(QMatrix))

\#

\# TCA map of rodent

tca.Data
$<$%
- tca(dataMatrix, nAxes=2,algorithm = "exhaustive")

plot(

tca.Data,

axes = c(1, 2),

labels.rc = c(1, 1),

col.rc = c("blue", "red"),

pch.rc = c(5, 5, 0.3, 0.3),

mass.rc = c(F, F),

cex.rc = c(0.6, 0.6),

jitter = c(F, T)

)

\#mfTCA map of rodent dimensions 1-2

tca.Data
$<$%
- tca(QMatrix, nAxes=2,algorithm = "exhaustive")

plot(

tca.Data,

axes = c(1, 2),

labels.rc = c(1, 1),

col.rc = c("blue", "red"),

pch.rc = c(5, 5, 0.3, 0.3),

mass.rc = c(F, T),

cex.rc = c(0.6, 0.6),

jitter = c(F, T)

)

\#mfTCA map of rodent dimensions 2-3

tca.Data
$<$%
- tca(QMatrix, nAxes=2,algorithm = "exhaustive")

plot(

tca.Data,

axes = c(2, 3),

labels.rc = c(1, 1),

col.rc = c("blue", "red"),

pch.rc = c(5, 5, 0.3, 0.3),

mass.rc = c(F, T),

cex.rc = c(0.6, 0.6),

jitter = c(F, T)

)

\section{Conclusion}

In his seminal paper Goodman (1996) introduced marginal-free correspondence
analysis; where his principal aim was to reconcile Pearson correlation measure
with Yule's association measure in the analysis of contingency tables. We
showed that marginal-free correspondence analysis is a particular case of
correspondence analysis with prespecified weights studied in the beginning of
the 1980s by Benz\'{e}cri and his students. Furthermore, we showed that it is
also a particular first-order approximation of logratio analysis with uniform
weights.\bigskip
\begin{verbatim}
Acknowledgements.
\end{verbatim}

Choulakian's research has been supported by NSERC of Canada. \bigskip

\textbf{References}

Aitchison J (1986) \textit{The Statistical Analysis of Compositional Data}.
London: Chapman and Hall.

Beh E and Lombardo R (2014) \textit{Correspondence Analysis: Theory, Practice
and New Strategies}. N.Y: Wiley.

Benz\'{e}cri JP (1973)\ \textit{L'Analyse des Donn\'{e}es: Vol. 2: L'Analyse
des Correspondances}. Paris: Dunod.

Benz\'{e}cri JP (1983a) Ajustement d'un tableau \`{a} des marges sous
l'hypoth\`{e}se d'absence d'interaction ternaire. \textit{Les Cahiers de
l'Analyse des Donn\'{e}es, }8\textit{(}2), 227-233

Benz\'{e}cri JP (1983b) Sur une g\'{e}n\'{e}ralisation du probl\`{e}me de
l'ajustement d'une mesure \`{a} des marges. \textit{Les cahiers de l'analyse
des donn\'{e}es}, 8(3), 359-370

Benz\'{e}cri JP, Bourgarit C, Madre JL (980) Probl\`{e}me: ajustement d'un
tableau \`{a} ses marges d'apr\`{e}s la formule de reconstitution. \textit{Les
Cahiers de l'Analyse des Donn\'{e}es}, 5(l), 163-172

Cholakian V (1980) Un exemple d'application de diverses m\'{e}thodes
d'ajustement d'un tableau \`{a} des marges impos\'{e}es. \textit{Les Cahiers
de l'Analyse des Donn\'{e}es,} 5(2), 173-176

Cholakian V (1984) M\'{e}thodes et crit\`{e}res pour l'ajustement d'un tableau
\`{a} des marges impos\'{e}es. \textit{Les Cahiers de l'Analyse des
Donn\'{e}es,} 9\textit{(}l), pp. 113-117

Choulakian V (2006) Taxicab correspondence analysis. \textit{Psychometrika,}
71, 333-345.

Choulakian V (2016) Matrix factorizations based on induced norms.
\textit{Statistics, Optimization and Information Computing}, 4, 1-14.

Choulakian V (2017) Taxicab correspondence analysis of sparse contingency
tables. \textit{Italian Journal of Applied Statistics,} 29 (2-3), 153-179.

Choulakian V (2021) Quantification of intrinsic quality of a principal
dimension in correspondence analysis and taxicab correspondence analysis.
Available on arXiv:2108.10685.

Cuadras CM, Cuadras D, Greenacre M (2006) A comparison of different methods
for representing categorical data. \textit{Communications in Statistics-Simul.
and Comp,} 35(2), 447-459.

Cuadras CM, Cuadras D (2015 ) A unified approach for the multivariate analysis
of contingency tables. \textit{Open Journal of Statistics}, 5, 223-232

Egozcue JJ, Pawlowsky-Glahn V, Templ M, Hron K (2015) Independence in
contingency tables using simplicial geometry. \textit{Communications in
Statistics - Theory and Methods}, 44:18, 3978-3996

Goodman LA (1979) Simple models for the analysis of association in
cross-classifications having ordered categories. \textit{Journal of the
American Statistical Association}, 74,537-55

Goodman LA (1981a) Association models and the bivariate normal for contingency
tables with ordered categories. \textit{Biometrika}, 68, 347-355

Goodman LA (1981b) Association models and canonical correlation in the
analysis of cross-classifications having ordered categories. \textit{Journal
of the American Statistical Association}, 76, 320-334

Goodman, LA (1991) Measures, models, and graphical displays in the analysis of
cross-classified data. \textit{Journal of the American Statistical
Association}, 86 (4), 1085-1111

Goodman LA (1996) A single general method for the analysis of cross-classified
data: Reconciliation and synthesis of some methods of Pearson, Yule, and
Fisher, and also some methods of correspondence analysis and association
analysis.\textit{\ Journal of the American Statistical Association}, 91, 408-428

Greenacre M (2009) Power transformations in correspondence analysis.
\textit{Computational Statistics \& Data Analysis,} 53(8), 3107-3116

Greenacre M (2010) Log-ratio analysis is a limiting case of correspondence
analysis. \textit{Mathematical Geosciences,} 42, 129-134

Madre JL (1980) M\'{e}thodes d'ajustement d'un tableau \`{a} des marges.
\textit{Les Cahiers de l'Analyse des Donn\'{e}es}, 5 (1), 87-99

Mosteller F (1968) Association and estimation in contingency tables,
\textit{Journal of the American Statistical Association}, 63, 1-28

Moussaoui AE (1987) Sur la reconstruction approch\'{e}e d'un tableau de
correspondance \`{a} partir du tableau cumul\'{e} par blocs suivant suivant
deux partitions des ensembles I et J. \textit{Les Cahiers de l'Analyse des
Donn\'{e}es, }12(3), 365-370

Yule, G.U. (1912). On the methods of measuring association between two
attributes. \textit{JRSS}, 75, 579-642

\bigskip
\begin{verbatim}
 
\end{verbatim}

\end{document}